\documentclass[aps,twocolumn]{revtex4}

\RequirePackage[T2A]{fontenc}

\usepackage{amssymb}
\usepackage{amsmath}
\usepackage{bm}
\usepackage[dvips]{epsfig}
\usepackage{graphicx}

\begin{document}

\title{Specificity of $\tau$ - approximation for chaotic electron 
trajectories on complex Fermi surfaces
}

\author{A.Ya. Maltsev}

\affiliation{
\centerline{\it L.D. Landau Institute for Theoretical Physics}
\centerline{\it 142432 Chernogolovka, pr. Ak. Semenova 1A,
maltsev@itp.ac.ru}}

\begin{abstract}
The work examines a special behavior of the magnetic conductivity 
of metals that arises when chaotic electron trajectories appear on the 
Fermi surface. This behavior is due to the scattering of electrons at 
singular points of the dynamic system describing the dynamics of 
electrons in $\, {\bf p}$ - space, and caused by small-angle scattering 
of electrons on phonons. In this situation, the electronic system 
is described by a ``non-standard'' relaxation time, which plays the 
main role in a certain range of temperature and magnetic field values.
\end{abstract}

\maketitle

\section{Introduction}

 In this work we will consider galvanomagnetic phenomena in pure metals 
in the limit of strong magnetic fields. This limit can be defined as 
the condition for a strong change of electronic states by the 
field during the electron free path time,  which is assumed to be 
sufficiently large. As was established in the 1950s - 1960s (in the 
school of I.M. Lifshitz), the key role in the description of 
galvanomagnetic phenomena in this limit is played by the geometry 
of semiclassical electron trajectories in presence of a magnetic field, 
determined by the system
\begin{equation} 
\label{MFSyst}
{\dot {\bf p}} \,\,\,\, = \,\,\,\, {e \over c} \,\,
\left[ {\bf v}_{\rm gr} ({\bf p}) \, \times \, {\bf B} \right]
\,\,\,\, = \,\,\,\, {e \over c} \,\, \left[ \nabla \epsilon ({\bf p})
\, \times \, {\bf B} \right] 
\end{equation}
(see \cite{lifazkag,lifpes1,lifpes2,etm}). 

 As is well known, the quantity $\, {\bf p} \, $ in the 
system (\ref{MFSyst}) is the quasi-momentum of a particle, determined 
up to the reciprocal lattice vectors. The system (\ref{MFSyst}) can be 
considered both as a system in the three-dimensional torus 
$\, \mathbb{T}^{3} = \mathbb{R}^{3} / L^{*} \, $ and as a system 
in the complete $\, {\bf p}$ - space $\, \mathbb{R}^{3} \, $. In the 
latter case, however, it is necessary to remember that the values 
of $\, {\bf p} \, $, differing by reciprocal lattice vectors, define 
the same quantum state. The dispersion relation 
$\, \epsilon ({\bf p}) \, $ can also be considered either as a smooth 
function on the torus $\, \mathbb{T}^{3} \, $, or as a 3-periodic 
function on $ \, \mathbb{R}^{3} $. At the same time, the motion 
of a particle in $\, {\bf x}$ - space is given by the relation
$${\dot {\bf x}} \,\,\,\, = \,\,\,\, {\bf v}_{\rm gr} ({\bf p})
\,\,\,\, = \,\,\,\, \nabla \epsilon ({\bf p}) $$

\vspace{1mm}

 From the point of view of system (\ref{MFSyst}), the condition of 
a strong magnetic field can be determined by the requirement that 
the electron covers a significant distance ($\gg p_{F}$) 
along the trajectories of this system between two successive 
scattering events. It is in this limit that most effects will be 
determined by the geometry of the trajectories of (\ref{MFSyst}), 
and the limit itself can be called the geometric limit.

 Formally, this limit can be written as the condition 
$\, \omega_{B} \tau \gg 1 \, $, where $\, \omega_{B} \, $ plays 
the role of the electron cyclotron frequency in metal, 
and $\, \tau\,$ represents the electron free path time.

\begin{figure}[t]
\begin{center}
\includegraphics[width=\linewidth]{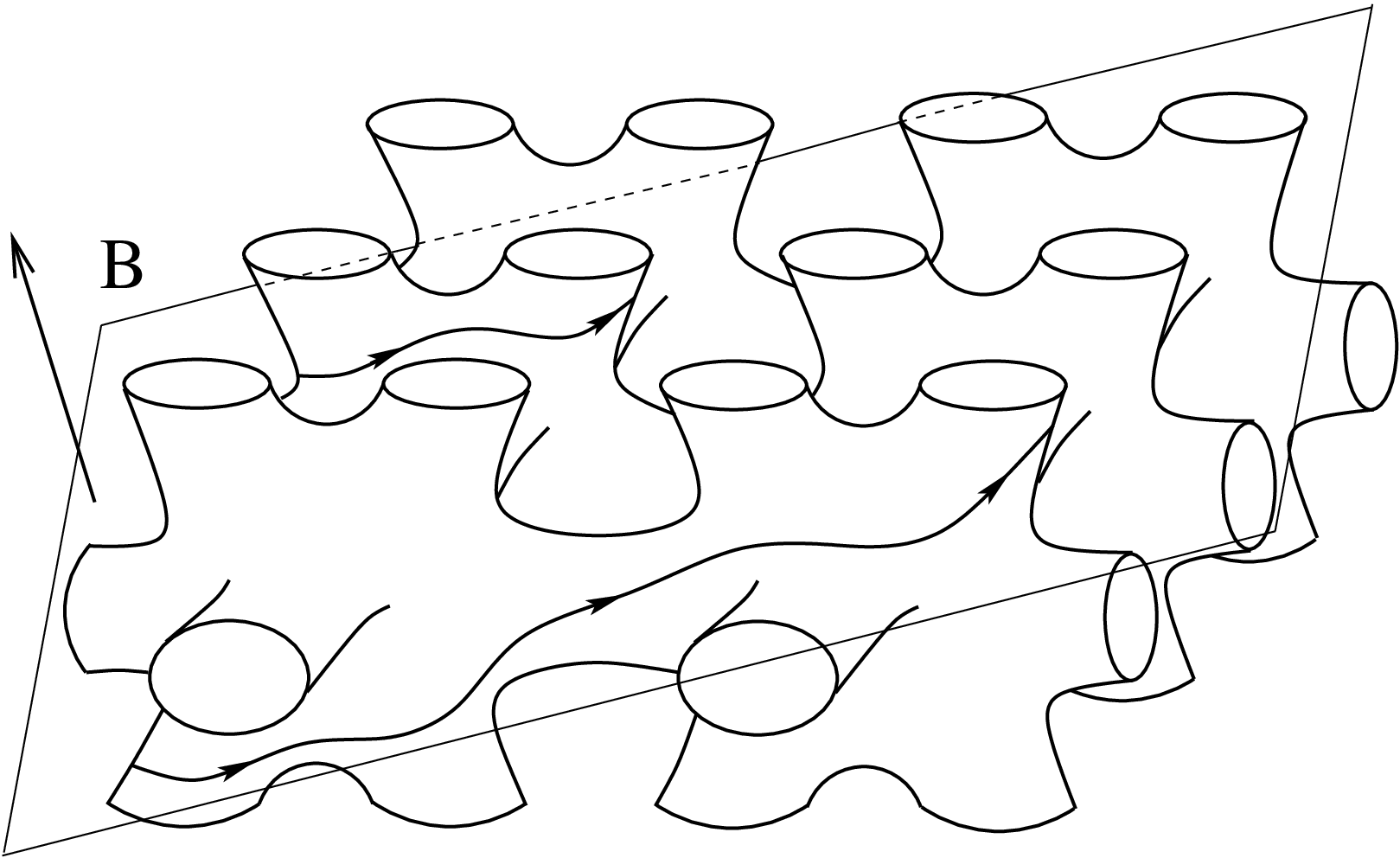}
\end{center}
\caption{Trajectories of system (\ref{MFSyst}) on a periodic 
Fermi surface of a rather complex shape}
\label{Fig1}
\end{figure}

 The trajectories of system (\ref{MFSyst}) in the 
full $\, {\bf p}$ - space are given by the intersections of 
planes orthogonal to $\, {\bf B} \, $ and the periodic surfaces 
$\, \epsilon ({\bf p}) = {\rm const} \, $ (Fig. \ref{Fig1}). 
As is also well known (see, for example, 
\cite{lifazkag,lifpes1,lifpes2,etm,Kittel,Abrikosov}), the main 
role in the physical effects is played by trajectories lying 
near the Fermi surface $\, \epsilon ({\bf p }) = \epsilon_{F} \, $.

 In describing galvanomagnetic phenomena (as well as other 
transport electronic phenomena), the greatest difference is 
observed between closed and open (unclosed) trajectories of 
(\ref{MFSyst}), which give significantly different contributions 
to transport phenomena in the limit $\, B \rightarrow \, \infty \, $. 
For example, one can see a huge difference in the contributions of 
closed and open periodic trajectories to the conductivity tensor 
in this limit (\cite{lifazkag})
\begin{equation}
\label{Closed}
\sigma^{kl} \,\,\,\, \simeq \,\,\,\,
{n e^{2} \tau \over m^{*}} \, \left(
\begin{array}{ccc}
( \omega_{B} \tau )^{-2}  &  ( \omega_{B} \tau )^{-1}  &
( \omega_{B} \tau )^{-1}  \cr
( \omega_{B} \tau )^{-1}  &  ( \omega_{B} \tau )^{-2}  &
( \omega_{B} \tau )^{-1}  \cr
( \omega_{B} \tau )^{-1}  &  ( \omega_{B} \tau )^{-1}  &  *
\end{array}  \right)  \quad  ,  
\end{equation}
$\omega_{B} \tau \, \rightarrow \, \infty $
(closed trajectories),
\begin{equation}
\label{Periodic}
\sigma^{kl} \,\,\,\, \simeq \,\,\,\,
{n e^{2} \tau \over m^{*}} \, \left(
\begin{array}{ccc}
( \omega_{B} \tau )^{-2}  &  ( \omega_{B} \tau )^{-1}  &
( \omega_{B} \tau )^{-1}  \cr
( \omega_{B} \tau )^{-1}  &  *  &  *  \cr
( \omega_{B} \tau )^{-1}  &  *  &  *
\end{array}  \right)  \quad  ,  
\end{equation}
$\omega_{B} \tau \, \rightarrow \, \infty $
(open periodic trajectories).

 In the formulas (\ref{Closed}) - (\ref{Periodic}), the 
quantity $\, n\, $ is of the order of the charge carrier 
concentration, and $\, m^{*}\, $ represents the effective 
electron mass in the crystal. In both cases, the direction 
of the $\, z\, $ axis coincides with the direction of the 
magnetic field; in addition, in the second case, the direction 
of the $\, x\, $ axis is chosen along the mean direction of 
periodic open trajectories in $\, {\bf p}$ - space. It can be 
seen that the main feature of the (\ref{Periodic}) mode is the 
strong anisotropy of conductivity in the plane orthogonal 
to ${\bf B}$, which obviously corresponds to the specific 
geometry of periodic open trajectories. Formulas 
(\ref{Closed}) - (\ref{Periodic}) determine the asymptotic 
behavior of the components of the conductivity tensor; 
in particular, all the given components contain, generally 
speaking, additional constant factors of order 1. For the 
value $\, \omega_ {B} \, $ we can use the approximate 
relation $\, \omega_{B} \simeq e B / m^{*} c \, $. In general, 
the above regimes are observed in fairly pure single crystals 
at fairly low temperatures and fairly large values of $\, B\, $. 
Both modes (\ref{Closed}) - (\ref{Periodic}) play an important 
role also in a more general case when trajectories of various 
shapes appear on the Fermi surface.

 At the same time, as was established later, the 
(\ref{Closed}) - (\ref{Periodic}) modes are not the only possible 
ones, and other types of open trajectories of system (\ref{MFSyst}) 
can give significantly different contributions to the magnetotransport 
phenomena in the limit $\, B \rightarrow \, \infty \, $.

\vspace{1mm} 

 The problem of classifying possible types of trajectories of 
system (\ref{MFSyst}) with an arbitrary dispersion relation was 
posed by S.P. Novikov in his work \cite{MultValAnMorseTheory}. 
This problem was then intensively studied in his topological school 
(see \cite{zorich1,dynn1992,Tsarev,dynn1,zorich2,DynnBuDA,dynn2,dynn3}) 
and can currently be considered solved in its main formulation. 
In particular, as a result of studies of the Novikov problem, 
all possible classes of open trajectories of the system (\ref{MFSyst}) 
were described, which can be divided into topologically regular 
(stable and unstable) and chaotic ones (of the Tsarev type and 
the Dynnikov type). Based on the mathematical results, it also 
became possible to introduce new topological quantities observed 
in the conductivity of normal metals 
(see \cite{PismaZhETF,UFN,BullBrazMathSoc,JournStatPhys}), as well as 
to describe new modes of behavior of the conductivity tensor in strong 
magnetic fields, which were unknown before (\cite{ZhETF1997,TrMian}).

 As can be seen from the formulas (\ref{Closed}) - (\ref{Periodic}), 
the contribution of both closed and periodic open trajectories to 
the conductivity tensor contains the parameter $\, \tau \, $, which 
plays the role of the relaxation time in the kinetic equation. This 
fact also occurs in more general cases, and, as we have already said, 
the geometric limit in magnetotransport phenomena corresponds to 
long relaxation times and fast dynamics of electronic states in a 
magnetic field. To calculate the main exponents in the asymptotic 
behavior of the conductivity tensor, it is convenient to use 
the $\, \tau$ - approximation in the kinetic equation which gives 
the correct laws for the decrease of the components 
$\, \sigma^{kl} (B) \, $ as $\, B \rightarrow \, \infty \, $.

  At the lowest temperatures, the time $\, \tau \, $ is 
determined mainly by the time of electron scattering on impurities 
$\, \tau_{imp} \, $. The intensity of electron-electron and 
electron-phonon scattering increases with increasing temperature, 
and at higher $T$ these processes become the main ones. When 
calculating conductivity, electron-phonon scattering processes 
appear later than anything else, which is caused by the small 
momentum of phonons at low $T$ and, as a consequence, long 
momentum relaxation times in these processes. As we will see below, 
however, in the most complex (from the ergodic point of view) cases, 
the above picture can change significantly. The reason for this is 
precisely electron-phonon scattering at small angles, which greatly 
changes the situation in the presence of trajectories with complex 
ergodic behavior. As a consequence of this, the role of 
electron-phonon collisions begins to manifest itself much earlier, 
and the most natural thing in this case is, in fact, the 
introduction of some effective value $\, \tau_{0} (B, T) \, $, 
determined not only by the scattering processes, but also by the 
features of the ergodic behavior of such trajectories.

 Here we will be interested in the most complex trajectories 
of system (\ref{MFSyst}), namely, the Dynnikov chaotic trajectories, 
which can arise only on Fermi surfaces of a sufficiently complex shape. 
The ergodic behavior of Dynnikov's trajectories is the most complex 
(both in planes orthogonal to ${\bf B}$ and on the Fermi surface 
$\, S_{F} $) and, as we will explain below, has important differences 
from the behavior of trajectories of other types. In particular, the 
asymptotic behavior of conductivity in the presence of trajectories 
of this type differs significantly from the (\ref{Closed}) and 
(\ref{Periodic}) regimes (see \cite{ZhETF1997,TrMian}).

 In the next section we will give a brief description of the 
general properties of Dynnikov’s chaotic trajectories, as well as 
the corresponding features of the conductivity tensor, which we need 
for further consideration. In section 3 we will consider the 
above-mentioned features of the relaxation time in strong magnetic 
fields, which, in fact, are inherent only in trajectories of this type.

\section{The emergence of chaotic trajectories and their geometric
properties}
\setcounter{equation}{0}

 Chaotic trajectories of system (\ref{MFSyst}) can arise only on 
sufficiently complex Fermi surfaces (see, for example, Fig. \ref{Fig1}) 
for specially selected directions of ${\bf B}$. In particular, the rank 
of the Fermi surface must be equal to 3, that is, the surface must extend 
in three independent directions in $\, {\bf p}$ - space.

  The behavior of trajectories of (\ref{MFSyst}) on complex Fermi 
surfaces depends quite complexly on the direction of the magnetic field. 
To describe them, it is convenient to use the angular diagram indicating 
the type of trajectories of (\ref{MFSyst}) for each of the directions 
of ${\bf B}$ (i.e. for each point on the unit sphere $\, \mathbb{S} ^{2} $).

 For almost every direction of ${\bf B}$, closed trajectories of the system 
(\ref{MFSyst}) are usually present on the Fermi surface. We can also 
separately identify the directions ${\bf B}$ for which only closed 
trajectories of (\ref{MFSyst}) are present on the Fermi surface. The 
corresponding directions ${\bf B}$ form open regions on the unit sphere, 
the union of which usually covers most of its area. Each of these 
regions can be attributed to the ``electronic'' or ``hole'' type, 
depending on whether the Hall conductivity is of the electronic or 
hole type for the corresponding directions of ${\bf B}$.

 Generic angular diagrams can be divided into two main types, namely, 
diagrams in which the regions described above correspond to the same 
type (electron or hole) and diagrams in which the regions of both types 
are present. We will call diagrams of the first type type A diagrams, 
and diagrams of the second type - type B diagrams.

 In addition to the regions corresponding to the presence of only closed 
trajectories, complex angular diagram contains directions ${\bf B}$ 
corresponding to the emergence of open trajectories of various types 
(periodic, topologically regular, Tsarev's type chaotic, Dynnikov's 
type chaotic) on the Fermi surface. As can be shown, for each of these 
directions ${\bf B}$ the emerging open trajectories are of the same type 
(see, for example, \cite{dynn3}). It is this circumstance that makes it 
especially convenient to use angular diagrams when describing trajectories 
of system (\ref{MFSyst}).
 
\begin{figure}[t]
\begin{center}
\includegraphics[width=\linewidth]{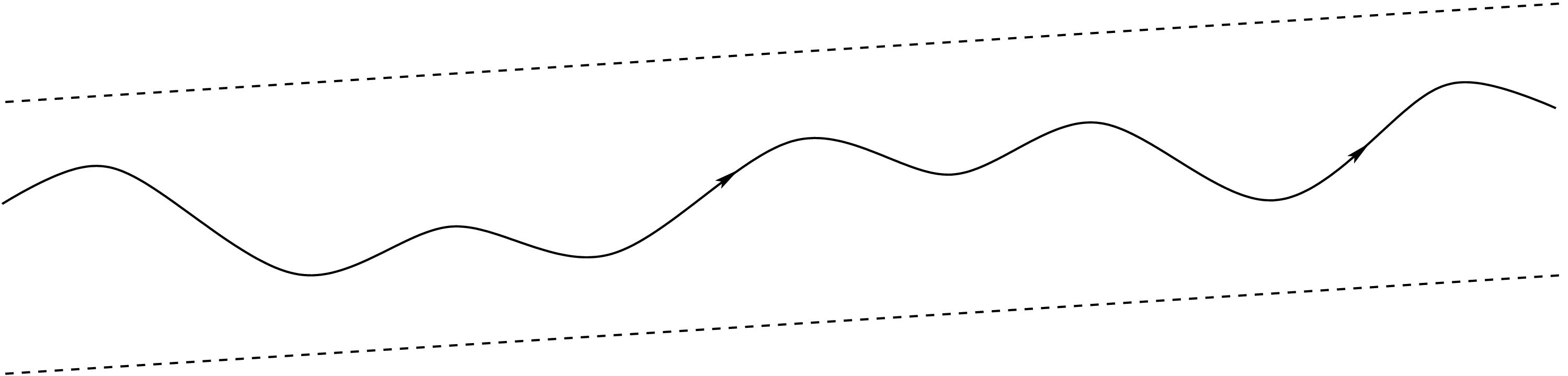}
\end{center}
\caption{Form of topologically regular open trajectories of 
system (\ref{MFSyst}) in planes orthogonal to ${\bf B}$.}
\label{Fig2}
\end{figure}

 The structure of complex angular diagrams is based on ``Stability Zones'' 
$\, \Omega_{\alpha} \subset \mathbb{S}^{2} \, $, corresponding to the 
emergence of ``topologically regular'' open trajectories of system 
(\ref {MFSyst}) on the Fermi surface. Topologically regular open 
trajectories are stable to all small rotations of ${\bf B}$ 
(as well as variations in the value of $\, \epsilon_{F}$) and have a 
relatively simple form in planes orthogonal to ${\bf B}$, namely, each 
such trajectory lies in a certain straight strip of finite width, passing 
through it (Fig. \ref{Fig2}).

 For the contribution of topologically regular trajectories 
to the tensor $\, \sigma^{kl} (B)\, $, in leading order, we can also use 
the formula (\ref{Periodic}), provided that the $\, x\, $ axis coincides 
with their mean direction in $\, {\bf p}$ - space.

 Each Stability Zone $\, \Omega_{\alpha} \, $ is a region with a piecewise 
smooth boundary on the sphere $ \, \mathbb{S}^{2} \, $ (see \cite{dynn3}) 
and corresponds, in fact, to some topological invariant observed in 
conductivity in strong magnetic fields (see \cite{PismaZhETF,UFN}).

\begin{figure}[t]
\begin{center}
\includegraphics[width=0.9\linewidth]{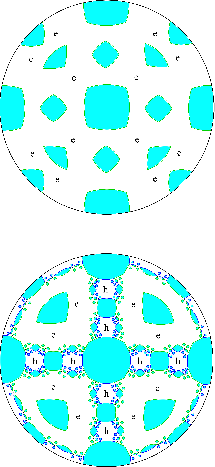}
\end{center}
\caption{Angular diagrams of type A (top) and B (bottom) (schematic). 
The letters ``e'' and ``h'' denote the sets of directions ${\bf B}$ 
corresponding to the presence of only closed trajectories on the Fermi 
surface and Hall conductivity of a fixed type (electron and hole, 
respectively).}
\label{Fig3}
\end{figure}

 An important difference between type A diagrams and type B diagrams 
is that type A diagrams contain a finite number of Stability Zones 
(Fig. \ref{Fig3}). In contrast, generic type B diagrams contain an 
infinite number of Zones $\, \Omega_{\alpha} \, $ 
(\cite{SecondBound,UltraCompl}). The zones $\, \Omega_{\alpha} \, $ 
in diagrams of type B form quasi-one-dimensional 
clusters on $ \, \mathbb{S}^{2} \, $, separating the regions of electron 
and hole Hall conductivity, corresponding to the presence of only closed 
trajectories on the Fermi surface (Fig. \ref{Fig3}).

 Clusters of Zones $\, \Omega_{\alpha} \, $ also contain infinite 
sets of directions ${\bf B}$ corresponding to the emergence of chaotic 
trajectories of (\ref{MFSyst}) on the Fermi surface. Thus, it is diagrams 
of type B that correspond to Fermi surfaces on which chaotic trajectories 
of system (\ref{MFSyst}) can appear. 

 Angular diagrams of type B correspond to the general situation and 
for generic dispersion relations $\, \epsilon ({\bf p}) \, $ arise in 
some finite energy interval 
$\, \epsilon_{F} \in ( \epsilon_{1} ^{\cal B} , \epsilon_{2}^{\cal B} ) $. 
At the same time, the width of the interval 
$\, ( \epsilon_{1}^{\cal B} , \epsilon_{2}^{\cal B} ) \, $ can be rather small 
for real relations $\, \epsilon ( {\bf p}) \, $. As a consequence, the search 
for materials satisfying the condition 
$\, \epsilon_{F} \in ( \epsilon_{1}^{\cal B} , \epsilon_{2}^{\cal B} ) $
represents a separate task. Here we will only note that for a number of 
materials, apparently, the emergence of a type B diagram can also be 
achieved by applying an external force to the sample (see \cite{PerLif}).

 As we have already said, chaotic trajectories of system (\ref{MFSyst}) 
are divided into two main classes, namely, Tsarev-type trajectories and 
Dynnikov-type trajectories. Tsarev-type trajectories have a simpler behavior 
in $\, {\bf p}$ - space and resemble topologically regular open trajectories 
(they have an asymptotic direction, however, cannot be contained in any 
straight strip of finite width in planes orthogonal to ${ \bf B}$). These 
trajectories, however, have a rather complex behavior on the compact Fermi 
surface $\, S_{F} \subset \mathbb{T}^{3} \, $ and according to this criterion 
they belong to the chaotic trajectories of the system (\ref{MFSyst}). We also 
note here that both Tsarev-type trajectories and Dynnikov-type trajectories 
are unstable with respect to small rotations of ${\bf B}$.

\begin{figure}[t]
\begin{center}
\includegraphics[width=\linewidth]{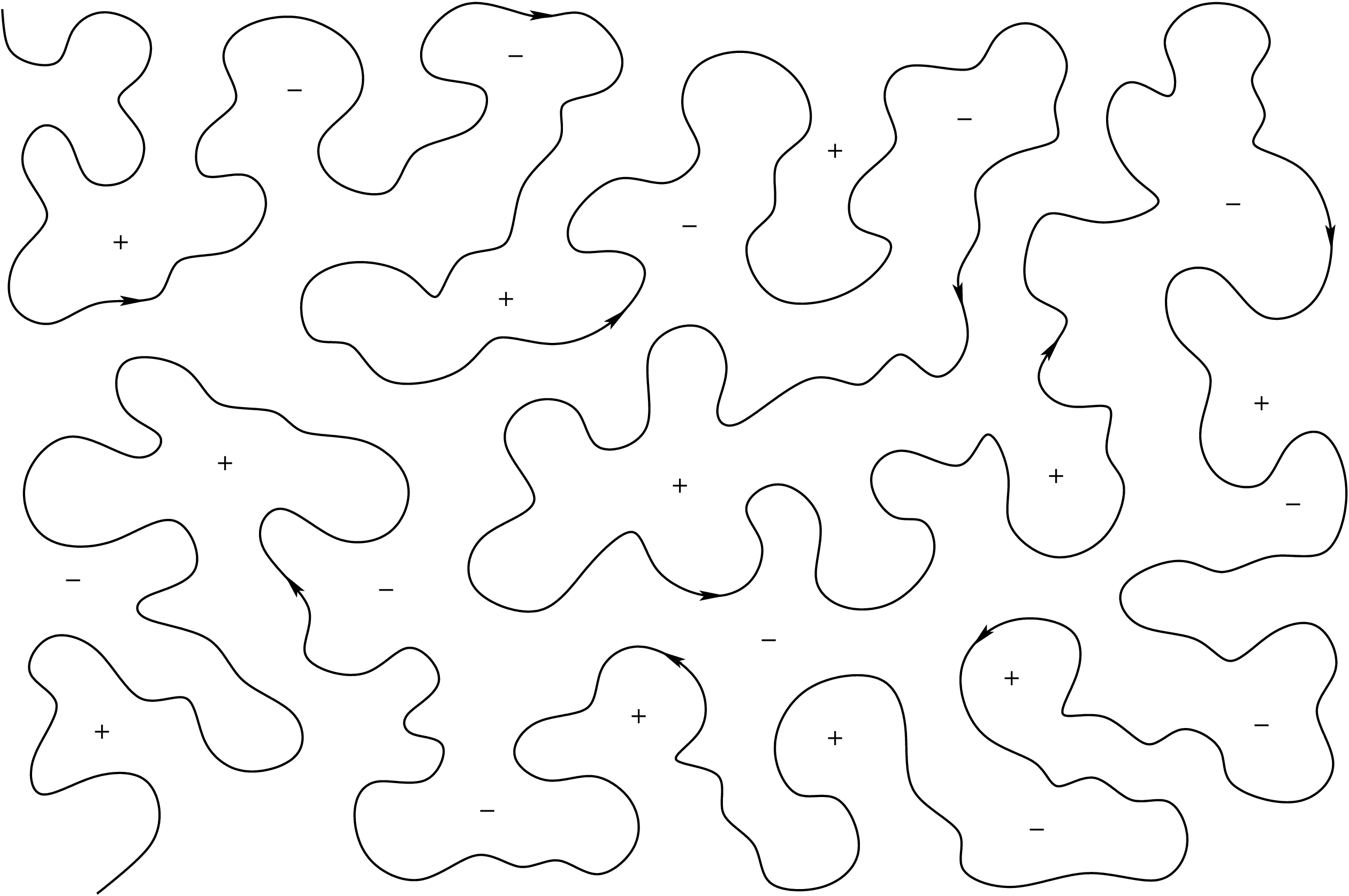}
\end{center}
\caption{The shape of Dynnikov’s chaotic trajectories in planes,
orthogonal to ${\bf B}$.}
\label{Fig4}
\end{figure}

 Trajectories of the Dynnikov type have the most complex behavior, 
exhibiting ``chaotic'' properties both in the full $\, {\bf p}$ - space 
and on the surface $\, S_{F} \subset \mathbb{T}^{ 3} \, $. In particular, 
such trajectories are characterized by ``chaotic'' wandering in planes 
orthogonal to ${\bf B}$, with gradual filling of all sections of such
planes (Fig. \ref{Fig4}).

\begin{figure}[t]
\begin{center}
\includegraphics[width=\linewidth]{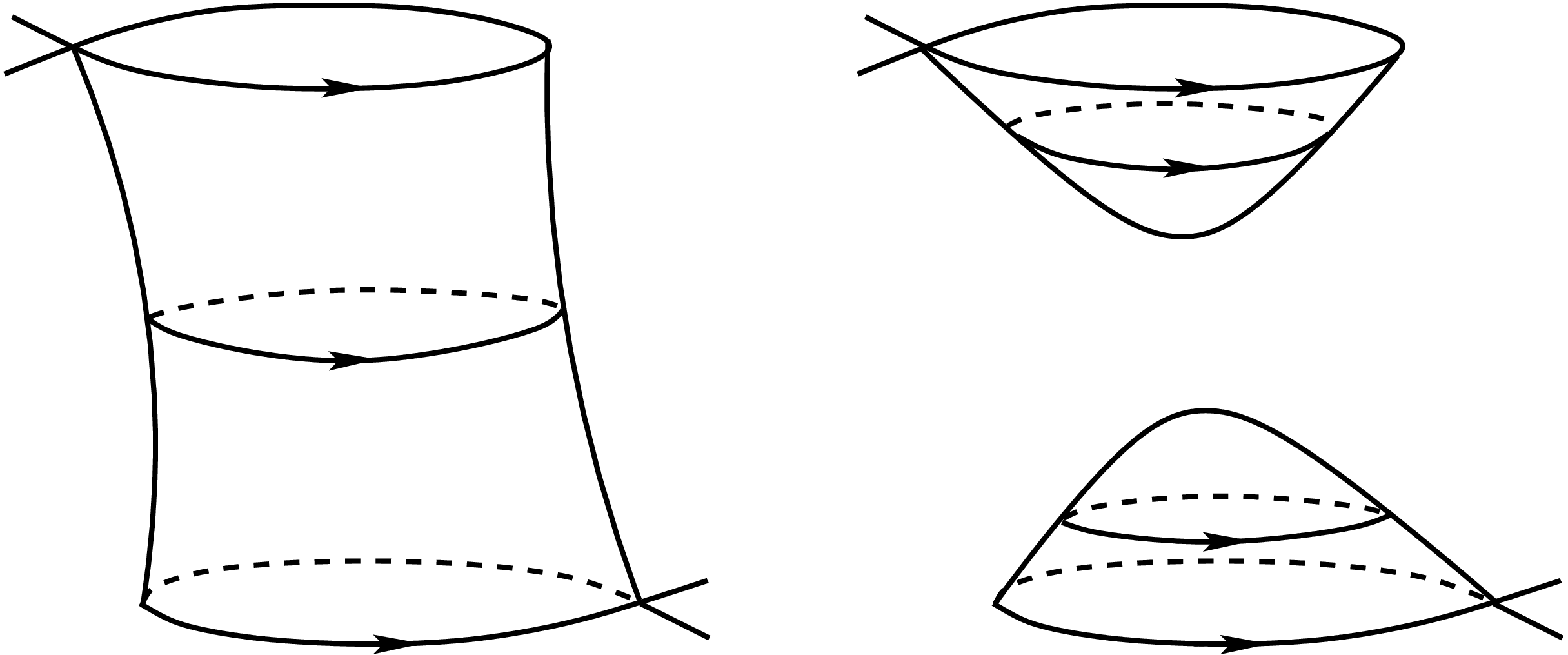}
\end{center}
\caption{Examples of cylinders of closed trajectories of 
system (\ref{MFSyst}) on the Fermi surface}
\label{Fig5}
\end{figure}

 When describing Dynnikov's trajectories on the surface 
$\, S_{F} \subset \mathbb{T}^{3} \, $ we must immediately mention 
that such trajectories, generally speaking, appear on the Fermi 
surface together with closed trajectories of the system (\ref {MFSyst}). 
The closed trajectories are combined into a finite number of 
(non-equivalent) cylinders bounded by singular closed trajectories of 
(\ref{MFSyst}) (Fig. \ref{Fig5}). Removing the cylinders of closed 
trajectories gives us a new surface (with edge) 
$\, {\widehat S}_{F} ({\bf B}) \, $ containing only open trajectories 
of system (\ref{MFSyst}).

 In the presence of Dynnikov's trajectories on the surface 
$\, S_{F} \, $, the surface $\, {\widehat S}_{F} ({\bf B}) \, $ 
remains a surface of rather high complexity, in particular, 
its genus (defined after filling the boundary of 
$\, {\widehat S}_{F} ({\bf B}) \, $ with flat disks) is always 
at least 3. In the overwhelming majority of cases, we can assume 
that $\, {\widehat S}_{F} ({\bf B}) \, $ is a surface of genus 3, 
invariant under the change $\, {\bf p} \rightarrow - {\bf p} \, $, 
and each chaotic trajectory is everywhere dense on the entire 
surface $\, {\widehat S}_{F} ({\bf B}) \, $. In general, the 
stochastic properties of Dynnikov’s trajectories have a huge 
number of very interesting features that are actively being 
studied at the present time (see, for example,
\cite{zorich2,DynnBuDA,dynn2,Zorich1996,ZorichAMS1997,
ZhETF1997,zorich3,DeLeo1,DeLeo2,ZorichLesHouches,DeLeoDynnikov1,
DeLeoDynnikov2,Dynnikov2008,Skripchenko1,Skripchenko2,DynnSkrip1,
DynnSkrip2,AvilaHubSkrip1,AvilaHubSkrip2,TrMian,DynHubSkrip}). 

 One of the consequences of such a complex behavior of Dynnikov's 
trajectories is their nontrivial contribution to the conductivity 
tensor in strong magnetic fields. In particular, this contribution 
vanishes in the limit $\, B \rightarrow \infty \, $ for all components 
$\, \sigma^{kl} (B) \, $, including conductivity along the direction 
of ${\bf B} $ (\cite{ZhETF1997}). In the interval 
$\, \omega_{B} \tau \gg 1 \, $ the components $\, \sigma^{kl} (B) \, $ 
have ``scaling'' behavior, reflecting the scaling properties of chaotic 
trajectories (\cite{ZhETF1997,TrMian}). It will also be especially 
important for us here that surfaces of this kind always contain saddle 
singular points of the system (\ref{MFSyst}), which have an important 
influence on the electron dynamics in the presence of a small-angle 
scattering.

 In the general case, the scaling behavior of chaotic trajectories has 
anisotropic properties and, with a suitable choice of the $\, x\, $ and 
$\, y\, $ axes, we can write for conductivity along the main directions
\begin{equation}
\label{sigmaxx}
\Delta \sigma^{xx} (B) \,\,\, \simeq \,\,\, 
{n e^{2} \tau \over m^{*}} 
\left( \omega_{B} \tau \right)^{2{\alpha}_{1} - 2} \,\,\, , 
\quad  \omega_{B} \tau \rightarrow \infty \,\,\,  , 
\end{equation}
\begin{equation}
\label{sigmayy}
\Delta \sigma^{yy} (B) \,\,\, \simeq \,\,\, 
{n e^{2} \tau \over m^{*}} 
\left( \omega_{B} \tau \right)^{2{\alpha}_{2} - 2} \,\,\, , 
\quad  \omega_{B} \tau \rightarrow \infty \,\,\,  , 
\end{equation}
\begin{equation}
\label{sigmazz}
\Delta \sigma^{zz} (B) \,\,\, \simeq \,\,\, 
{n e^{2} \tau \over m^{*}} 
\left( \omega_{B} \tau \right)^{2{\alpha}_{3} - 2} \,\,\, , 
\quad  \omega_{B} \tau \rightarrow \infty 
\end{equation}
($0 < \alpha_{1}, \alpha_{2}, \alpha_{3} < 1$). 

\begin{figure}[t]
\begin{center}
\includegraphics[width=\linewidth]{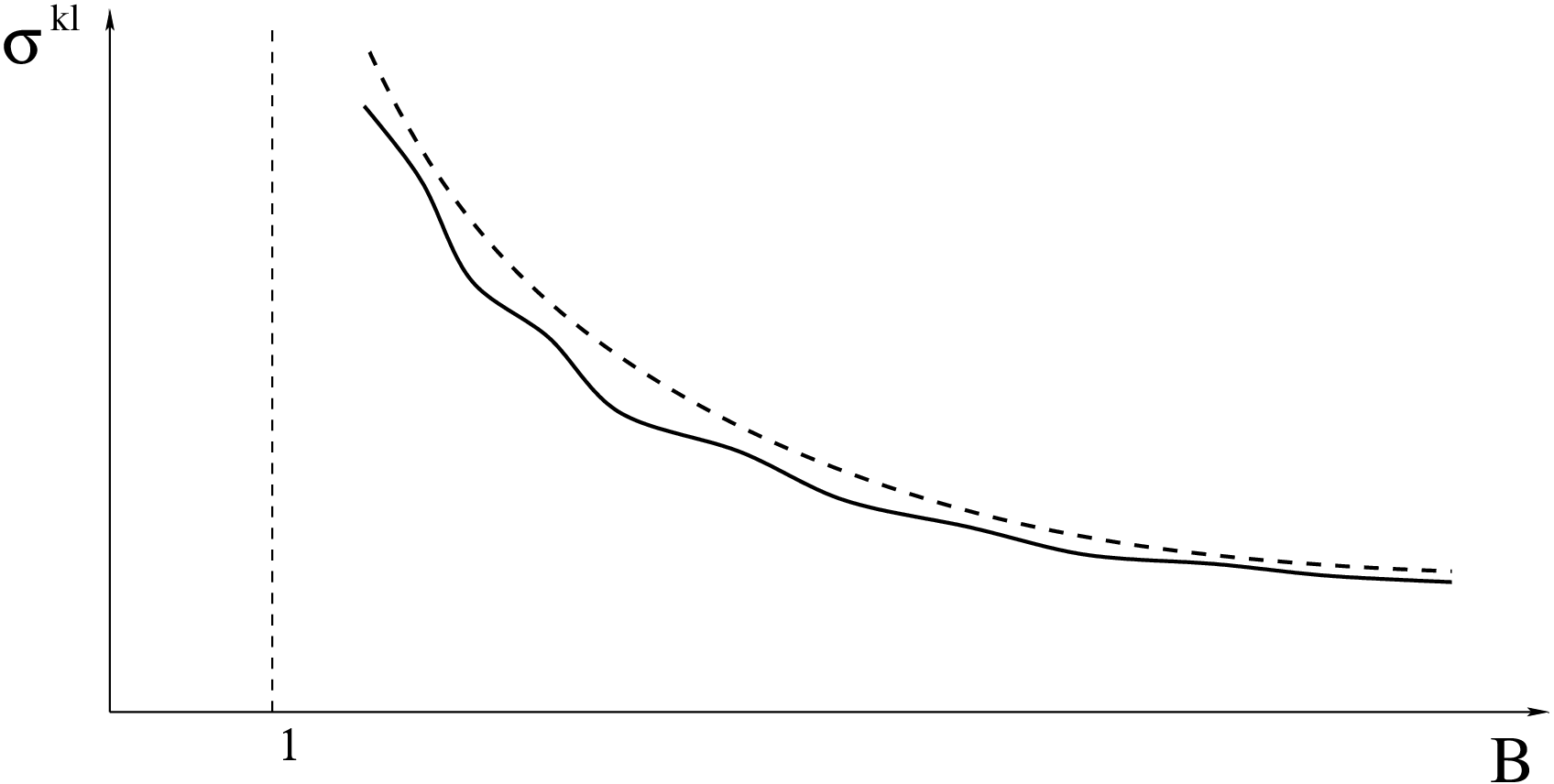}
\end{center}
\caption{Behavior of the components of the conductivity 
tensor corresponding to the contribution of chaotic trajectories 
of (\ref{MFSyst}) of Dynnikov's type}
\label{Fig6}
\end{figure}

 Here we should immediately note that, in contrast to the 
relations (\ref{Closed}) - (\ref{Periodic}), the relations 
(\ref{sigmaxx}) - (\ref{sigmazz}) are not the main term of 
any asymptotic expansion for $\, \sigma^{kl} (B) \, $. Instead, 
they define a general ``trend'' of decreasing components 
$\, \sigma^{kl} (B) \, $ as $\, B \rightarrow \infty \, $, 
which may also have an additional (cascade) structure in the 
interval $\, \omega_{B} \tau \gg 1 \, $ (Fig. \ref{Fig6}).

  It can also be noted that the contribution of chaotic 
trajectories to conductivity should in general be added 
to the contribution of closed trajectories (\ref{Closed}), 
which can also be present on the Fermi surface. It can be 
seen that the contribution of chaotic trajectories noticeably 
exceeds the contribution of closed trajectories to conductivity 
in the plane orthogonal to ${\bf B}$, and is noticeably less than 
their contribution to conductivity along the magnetic field. From 
this point of view, perhaps it is the study of conductivity in the 
plane orthogonal to ${\bf B}$ that is most convenient when studying 
the geometry of chaotic trajectories.

  As a rule, when studying the geometric properties of trajectories 
of (\ref{MFSyst}), the dependence of $\, \sigma^{kl} (B) \, $ on the 
value of $B$ is studied for a fixed (maximum) value of $\tau$. Here 
we will be interested in its dependence on both quantities 
$B$ and $\tau$ in the interval $\, \omega_{B} \tau \gg 1 \, $.

 As can be seen from the formulas (\ref{sigmaxx}) - (\ref{sigmazz}), 
the values $\, \sigma^{ll} (B)\, $ can both decrease 
($\alpha_{l} < 1/ 2$), and increase ($\alpha_{l} > 1/2$), with 
increasing $\tau$. This circumstance, in fact, is caused by an increase 
in the values of $\, \sigma^{ll} \, $ for $\, B = 0 \, $ with 
increasing $\tau$, and the ratio 
$\, \sigma^{ll} (B, \tau) / \sigma^ {ll} (0, \tau) \, $ 
is a decreasing function of $\tau$. Note also that the relations 
(\ref{sigmaxx}) - (\ref{sigmazz}) should be used in the interval 
$\, \sigma^{ll} (B, \tau) \ll \sigma^{ll} (0, \tau) \, $ 
($\omega_{B} \tau \gg 1$).

 In general, both dependences of $\, \sigma^{ll} (B, \tau) \, $ 
on both arguments can be used to determine the scaling parameters 
of chaotic trajectories. The dependence on $\tau$ (for a fixed $B$) 
is strongest for $\, \alpha_{l} \, $, noticeably different 
from $\, 1/2 \, $, and disappears for $\, \alpha_{l} \simeq 1/2 \, $.

 A more detailed discussion of the relations 
(\ref{sigmaxx}) - (\ref{sigmazz}) is presented in \cite{PismaZhETF,UFN}. 
Here we note only the main reason for this behavior of the components 
$\, \sigma^{kl} (B) \, $. As we have already said, it lies in the 
geometric properties of the chaotic trajectories in $\, {\bf p}$ - space,
as well as their properties on the compact Fermi surface 
$\, S_{F} \subset \mathbb{T}^{3 } \, $.

 Namely, for the behavior of conductivity (and other magnetotransport 
phenomena), the geometry of sections of chaotic trajectories of length 
of the order of $\, v_{F} \tau \, $ in coordinate space 
(or $\, l \sim p_{F} \omega_{B} \tau \, $ in $\, {\bf p}$ - space)
turns out to be especially important. More precisely, it is important 
to know the average deviation of the ends of such sections along each 
of the coordinates $\, x\, $, $\, y\, $ and $\, z\, $. The corresponding 
averages grow in a power-law manner in $\, {\bf p}$ - space
$$| \Delta p_{x} (l) | \,\, \simeq \,\, p_{F} 
\left( {l \over p_{F}} \right)^{{\alpha}_{2}} \,\,\, , \quad 
| \Delta p_{y} (l) | \,\, \simeq \,\, p_{F} 
\left( {l \over p_{F}} \right)^{{\alpha}_{1}} $$
and  respectively,
$$| \Delta x^{i} | \,\,\, \sim \,\,\, {v_{F} \over \omega_{B}} \, 
\left( \omega_{B} \tau \right)^{\alpha_{i}} 
\,\,\, \sim \,\,\, {c p_{F} \over e B} \, 
\left( \omega_{B} \tau \right)^{\alpha_{i}} $$
in coordinate space.

 The values $\, \alpha_{i} \, $ lie in the interval 
$\, (0, 1) \, $, and we have different degrees 
$\, \alpha_{2} \, $ and $\, \alpha_{1} \, $ for some principal 
directions $\, p_{x} \, $ and $\, p_{y} \, $ in 
$\, {\bf p}$ - space. This behavior also extends to coordinate 
space (note that the projections of trajectories in 
$\, {\bf x}$ - space onto the plane orthogonal to ${\bf B}$ 
are similar to the trajectories in $\, {\bf p}$ - space rotated 
by $90^{\circ}$). Separately, the scaling parameter 
$\, \alpha_{3} \, $ arises for deviations along 
the $\, z \, $ axis in the coordinate space. Anistropic 
scaling behavior of the quantities $\, | \Delta x^{i} | \, $ 
is expressed in the corresponding anisotropy of the electron 
drift in an external electric field, which, in turn, is expressed 
in the dependences (\ref{sigmaxx}) - (\ref{sigmazz}).

 From the kinetic equation in the $\, \tau$ - approximation it is 
not difficult to get the formula
$$\Delta \, s^{kl}(B) \,\,\,\, = \,\,\,\,
e^{2} \, \tau \,\, \iint_{\widehat{S}_{F}} \,
\langle v^{k}_{\rm gr} \rangle_{B} \,\,
\langle v^{l}_{\rm gr} \rangle_{B}
\,\,\, {d p_{z} \, d s \over (2 \pi \hbar)^{3}}  $$
for the contribution of chaotic trajectories to the symmetric part 
of the conductivity tensor (taking into account spin), 
where $\, s = t e B / c \, $ and $\, t\, $ is the travel time along 
the trajectories.

 The values $\, \langle v^{k}_{\rm gr} \rangle_{B} \, (p_{z}, t) \, $ 
are defined by the averaging on the corresponding trajectory
$$\langle v^{k}_{\rm gr} \rangle_{B} \, (p_{z}, t)
\,\,\,\, \equiv \,\,\,\,
{1 \over \tau} \, \int_{-\infty}^{t} \,
v^{k}_{\rm gr} \,(p_{z}, t^{\prime}) \,\,\,
e^{{(t^{\prime} - t) \over \tau}} \,\, d t^{\prime} \,\,\, , $$
and can be approximated by the formula
$$\langle v^{k}_{\rm gr} \rangle_{B} \, (p_{z}, s)
\,\,\, \simeq \,\,\,  {1 \over \tau} \,
\int_{t - \tau}^{t} \,\,\,
v^{k}_{\rm gr} \,(p_{z}, t^{\prime}) \,\, d t^{\prime} $$
for large values of $ \tau$.

 Assuming directly from the system (\ref{MFSyst}):
$$| \langle v^{x}_{\rm gr} \rangle_{B} | \,\, = \,\,
\left| { c \, \Delta p_{y} \over e B \tau } \,  \right| \,\,\, ,  \quad 
| \langle v^{y}_{\rm gr} \rangle_{B} | \,\, = \,\,
\left| { c \, \Delta p_{x} \over e B \tau } \,  \right| \,\, ,  $$
we in the same way obtain a connection between the scaling 
parameters of the trajectory in $\, {\bf p}$ - space with 
the scaling parameters of the conductivity tensor. (Similarly, 
the scaling parameter $\, \alpha_{3} \, $ arises from the estimating 
of the average value $\, \langle v^{z}_{\rm gr} \rangle_{B} \, $ 
along the trajectory).

\vspace{1mm}

 It can be seen that the behavior of Dynnikov's trajectories is
noticeably different from ordinary diffusion, despite their 
obvious ``chaotic'' wandering in planes orthogonal to ${\bf B}$. 
To some extent, this is explained by the absence of self-intersections 
in such trajectories, and in general, by the presence of velocity 
correlations (with non-trivial scaling properties) on all their scales.

 It is extremely important in our situation that the main directions, 
as well as the scaling parameters $\, \alpha_{i} \, $ 
(the Zorich - Kontsevich - Forney indices) are the same for all 
chaotic trajectories (in all planes orthogonal to magnetic field)
for a given direction of $\, {\bf B} \, $. This is due to a specific 
behavior of such trajectories on the surface 
$\, {\widehat S}_{F} ({\bf B})\, $, reflecting the general features 
of the system (\ref{MFSyst}).

\begin{figure}[t]
\begin{center}
\includegraphics[width=\linewidth]{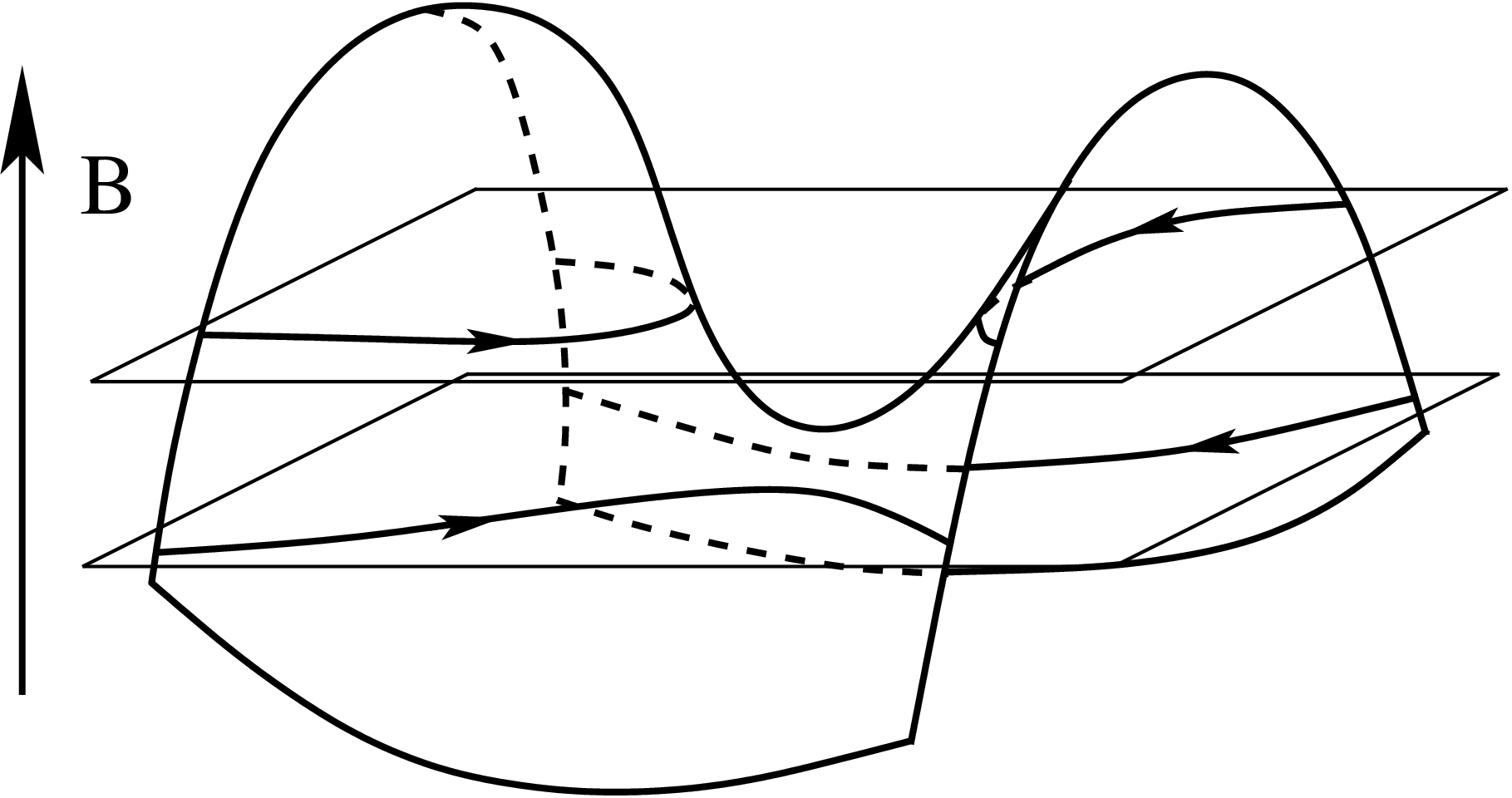}
\end{center}
\caption{A reconstruction of a chaotic trajectory in 
$\, {\bf p}$ - space during a crossing of a saddle singular 
point by a plane orthogonal to ${\bf B}$}
\label{Fig7}
\end{figure}

\begin{figure}[t]
\begin{center}
\includegraphics[width=\linewidth]{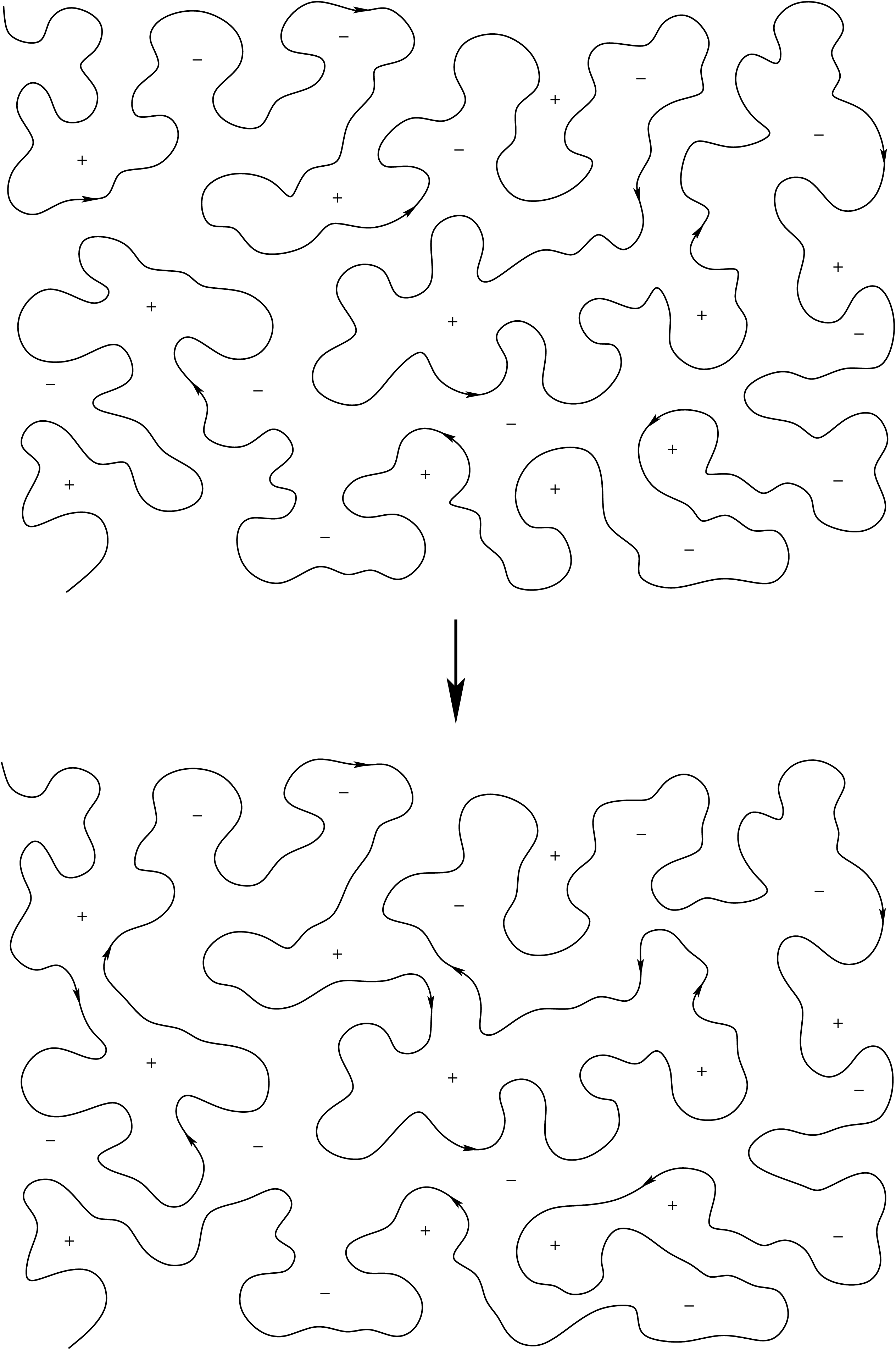}
\end{center}
\caption{Multiple reconstructions of a chaotic trajectory 
in $\, {\bf p}$ - space when changing the value of $\, p_{z}\, $ 
(schematically)}
\label{Fig8}
\end{figure}

 The above properties hold in all planes orthogonal to ${\bf B}$, 
despite multiple reconstructions of chaotic trajectories with changing 
$\, p_{z}\, $. The latter occur due to the presence of saddle singular 
points of the system (\ref{MFSyst}) inside the surface 
$\, {\widehat S}_{F} ({\bf B}) \, $, which cause such reconstructions 
when they are intersected by planes orthogonal to $ {\bf B}$ 
(Fig. \ref{Fig7}). These points, repeating periodically in 
$\, {\bf p}$ - space, cause reconstructions in the geometry of 
chaotic trajectories on all scales (Fig. \ref{Fig8}). At the same time, 
as we have already said, this does not change the main directions and 
scaling parameters of the trajectories. The last property is explained 
by the fact that reconstructions at different points are not independent, 
but, in fact, are coordinated with each other in a special (complex) way.

 The parameter $\, \tau \, $, as is easy to see, plays the role of 
the time of destruction of correlations in the motion of a particle 
in $\, {\bf x} $ - $\, $ and $\, {\bf p}$ - space. The parameter 
$\, \omega_{B} \tau \, $ determines the scale of the geometric length 
(in $\, {\bf p}$ - space) at which correlations are still preserved. 
Looking ahead, we note that here we will be interested in the processes 
of destruction of correlations in the motion of an electron, caused by 
the ``vibration'' of the value $\, p_{z} \, $ due to small-angle 
scattering on phonons, leading to random reconstructions of chaotic 
trajectories near singular points of the system (\ref{MFSyst}). 
These processes, as we will see, lead to the emergence of a new 
effective value $\, \tau \, $, which in general differs from the 
relaxation times due to other scattering processes.

\section{Mean free path time on chaotic trajectories}
\setcounter{equation}{0}

 We will now consider the behavior of the free path time 
of electrons on chaotic trajectories in the low temperature 
regime, where this time is quite large. As is well known 
(see, for example, \cite{etm,Kittel,Abrikosov}), the free 
path time of electrons in a single crystal is determined 
mainly by three processes, namely electron-electron scattering, 
electron-phonon scattering and scattering by impurities. 
The first two processes are characterized by a strong 
dependence on temperature, while the latter is almost 
independent of $\, T $.

 To achieve the longest free path time, very clean samples 
are usually used at very low temperatures. In this case, 
electron-electron and electron-phonon scattering become 
insignificant, and the time $\, \tau \, $ is completely 
determined by residual scattering on impurities. In our case, 
however, we will consider slightly higher temperatures, 
when all three of these processes appear. 

 Let us present here the most approximate estimates of the 
temperature intervals corresponding to the intensities of 
the above processes that interest us, using the most general 
assumptions.

 Let us note first of all that in the purest samples the time 
of electron scattering on impurities $\, \tau_{imp} \, $ can reach 
$10^{-8} \, {\rm sec}$, which corresponds to the mean free path 
$\, l \sim 1 \, {\rm cm} \, $. This value can apparently be used 
as an upper estimate for $\, \tau_{imp} \, $, although, in reality, 
for our effects, noticeably smaller values of $\, \tau \, $ are 
often sufficient (note, that for $\, \tau \sim 10^{-9} \, {\rm sec} \, $ 
the value of $\, \omega_{B} \tau \, $ in the interval 
$\, 0.1 \, {\rm Tl} \leq B \leq 10 \, {\rm Tl} \, $ can be estimated 
as $\, 10 \leq \omega_{B} \tau \leq 1000 \, $, which is, of course, 
enough to manifest the geometry of complex trajectories we are 
considering here). 

 The time of the electron-electron scattering can be estimated 
(see, for example, \cite{Abrikosov}) using the formula
\begin{equation}
\label{ee}
\tau_{ee} \,\,\, \simeq \,\,\, {\hbar \over k T} \,
{\epsilon_{F} \over k T} 
\end{equation}
(where $\, \hbar \simeq 6.582 \cdot 10^{-16} \, {\rm eV} \cdot {\rm sec} \, $,
$\, \epsilon_{F} \simeq 5 \, {\rm eV} $), which gives the value
$\, \tau_{ee} \simeq 10^{-8} \, {\rm sec} \, $ at $\, T = 5 \, {\rm K} \, $. 

 Thus, (roughly) we can assume that in our situation the processes 
of electron-electron scattering become insignificant in comparison 
with the processes of scattering by impurities already at temperatures 
of the order of several kelvins.

 The average time between electron scatterings on phonons can be 
(see \cite{Abrikosov}) estimated as
\begin{equation}
\label{ephepsilon}
\tau^{\epsilon}_{ph} \,\,\, \simeq \,\,\, {\hbar \over k T} \,
\left( {T_{D} \over T} \right)^{2}
\end{equation}

 As is well known, however, this time is the time of the energy 
relaxation of electrons due to scattering by phonons. The momentum 
relaxation time $\, \tau^{p}_{ph} \, $ is much longer in this case, 
since the phonons have a very small momentum 
$$p_{ph} \,\,\, \simeq \,\,\, {T \over T_{D}} \, p_{F} 
\,\,\, \ll \,\,\, p_{F} \,\,\, , $$
transferred to electrons during the scattering processes. As a result, 
an electron undergoes a ``diffusion'' motion along the Fermi surface, 
repeatedly scattering on phonons, and its momentum relaxation 
time can be estimated as
$$\tau^{p}_{ph} \,\,\, \simeq \,\,\, \tau^{e}_{ph} \, 
\left( {T_{D} \over T} \right)^{2} \,\,\, \simeq \,\,\,
{\hbar \over k T} \, \left( {T_{D} \over T} \right)^{4} $$

  Setting $\, T_{D} \simeq 300 \, {\rm K} \, $, for the value 
$\, T = 5 \, {\rm K} \, $ we obtain the values 
$\, \tau^{ \epsilon}_{ph} \simeq 5 \cdot 10^{-9} \, {\rm sec} \, $ 
and $\, \tau^{p}_{ph} \simeq 2 \cdot 10^{ -5} \, {\rm sec} \, $. 
It can be seen, therefore, that for given parameters, the 
temperature region where all three relaxation processes are 
significant lies near the values of $\, T \simeq 5 \, {\rm K} \, $. 
Certainly, all the above estimates are correct only in order of 
magnitude; in addition, the values $\, \epsilon_{F} \, $ and 
$\, T_{D} \, $ can differ noticeably (by an order of magnitude) 
for different substances. In general, we can assume that the 
interval of interest to us lies near the values of $\, T\, $ of 
the order of ten (or tens) kelvins and strongly depends on the 
individual parameters of a conductor.

 As can also be seen, in the above example, the electron-phonon 
scattering does not play a big role in calculating conductivity 
in the standard situation due to the large value of $\,\tau^{p}_{ph}\,$ 
compared to other times. (At the same time, it plays the main role 
in calculating thermal conductivity due to the small value 
of $\, \tau^{\epsilon}_{ph}$).

\vspace{1mm}

 As we said above, we are going to consider here a ``non-standard'' 
situation, namely, the situation of the emergence of chaotic 
trajectories on the Fermi surface. As we have also already said, 
the main role in this case will be played by the presence of saddle 
singular points inside the carrier of chaotic trajectories. The 
source of the special behavior of $\tau$ in this case is 
the ``vibration'' of the value of $\, p_{z} \, $ due to small-angle 
scattering on phonons, leading to ambiguity of motion along a 
trajectory near singular points (Fig. \ref {Fig7}). In this situation, 
small-angle scattering begins to play a very significant role and, 
thus, the time $\, \tau_{ph}^{\epsilon} \, $ becomes significant 
when calculating $\, \sigma^{kl} (B) \, $ in strong magnetic fields.

 We will be most interested in the situation when this effect is 
the main one, which implies the relations
$$\tau_{ph}^{\epsilon} \,\,\, < \,\,\, \tau_{ee} \, , \, \tau_{imp} $$

 As we saw above, for ``standard'' values $\, \epsilon_{F} \, $ 
(5 {\rm eV}) and $\, T_{D} \, $ (300 {\rm K}) in ultrapure metals, 
this relation holds for $\, T \geq 5 \, {\rm K} \, $. This threshold 
value may be lower for materials with special parameters (relatively 
speaking, high values of $\, \epsilon_{F} \, $ and low values 
of $\, T_{D} \, $, as well as weaker electron-electron interaction 
and strong electron-phonon interaction).

 The temperature range suitable for us is limited from above by 
the condition $\, \omega_{B} \tau_{ph}^{\epsilon} \gg 1 \, $. Assuming, 
for example, $\, \omega_{B} \tau_{ph}^{\epsilon} \geq 10 \, $, for 
the ``standard'' value $\, T_{D} \simeq 300 \, {\rm K} \, $ we get 
the estimates
$$T \, \leq \, 40 \, {\rm K} \quad \text{at} \quad 
B \, \simeq \, 10 \, {\rm Tl} \,\,\, , $$
$$T \, \leq \, 20 \, {\rm K} \quad \text{at} \quad 
B \, \simeq \, 1 \, {\rm Tl} $$
(The upper threshold value of $T$ may be higher for materials 
with larger values of $T_{D}$, as well as for larger values of $B$).

\vspace{1mm}

 As we have already said, in our situation another free path 
time $\, \tau_{0} (B, T) \, $ arises, which is determined by 
the ``scattering'' of electrons at the saddles of the 
system (\ref{MFSyst}). 

 To estimate the time $\, \tau_{0} (B, T) \, $ we must consider 
changes in the value of $\, p_{z} \, $ along the chaotic 
trajectories of system (\ref{MFSyst}). Since such changes 
are caused by small-angle scattering on phonons, we, in any case, 
have the inequality
$$\tau_{0} (B, T) \,\,\, \geq \,\,\, \tau_{ph}^{\epsilon} $$

 The change in electron momentum during each scattering is 
of the order of
$$\delta p_{0} \,\,\, \simeq \,\,\, {T \over T_{D}} \, p_{F}
\,\,\, \ll \,\,\, p_{F} $$ 

 The change in electron energy in this case is of the order of
$\, \delta \epsilon \, \simeq \, k T \, \ll \, \epsilon_{F} \, $.
Note that due to the relation
$$k T / \epsilon_{F} \,\,\, \ll \,\,\, T / T_{D} $$
we can assume that the electron actually remains on the Fermi 
surface all the time, ``drifting'' along the trajectories of 
system (\ref{MFSyst}) with multiple scattering on phonons.

\begin{figure}[t]
\begin{center}
\includegraphics[width=\linewidth]{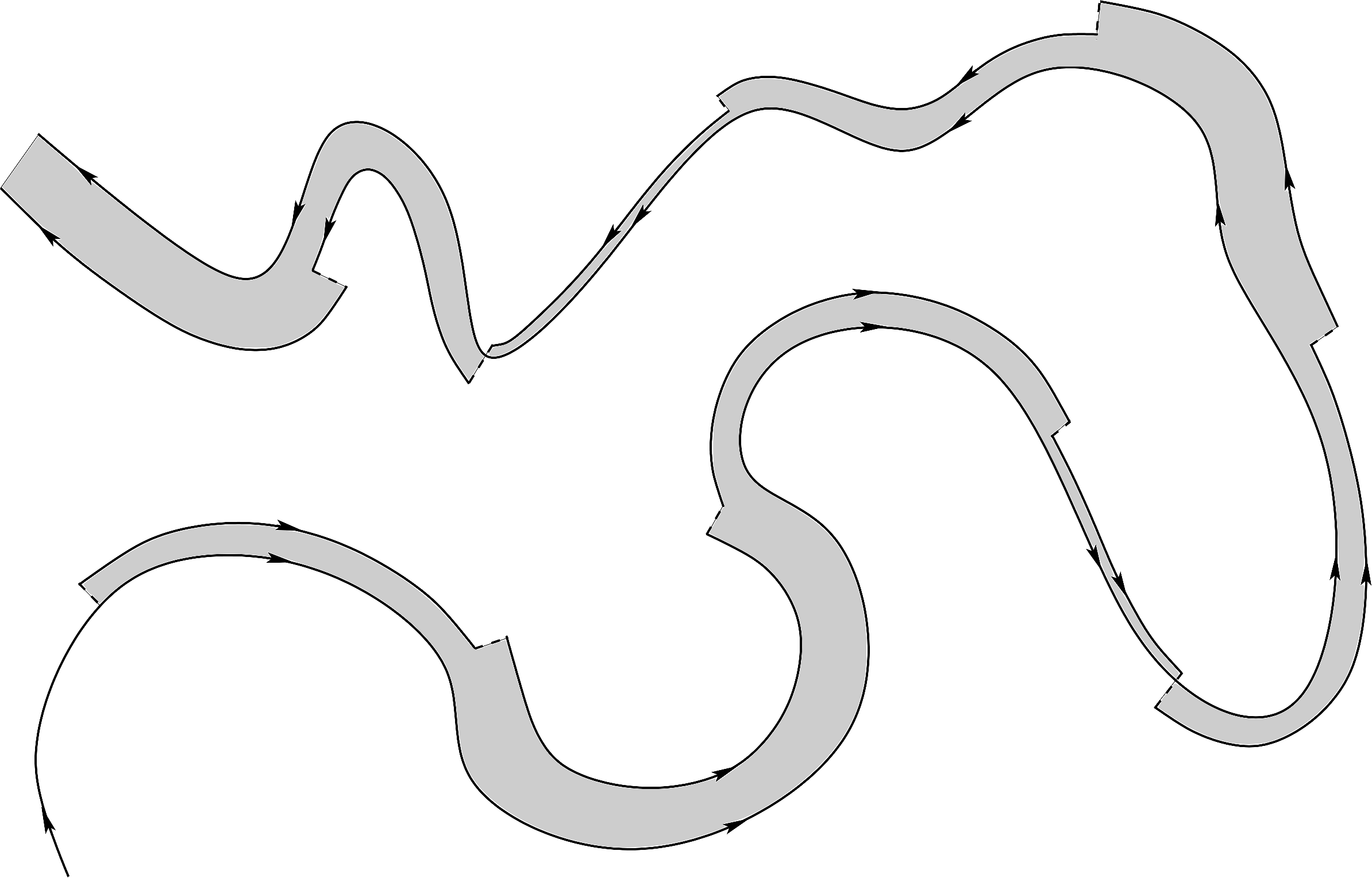}
\end{center}
\caption{Narrow band limited by the original and true electron 
trajectory on a complex Fermi surface (schematically)}
\label{Fig9}
\end{figure}

\begin{figure}[t]
\begin{center}
\includegraphics[width=\linewidth]{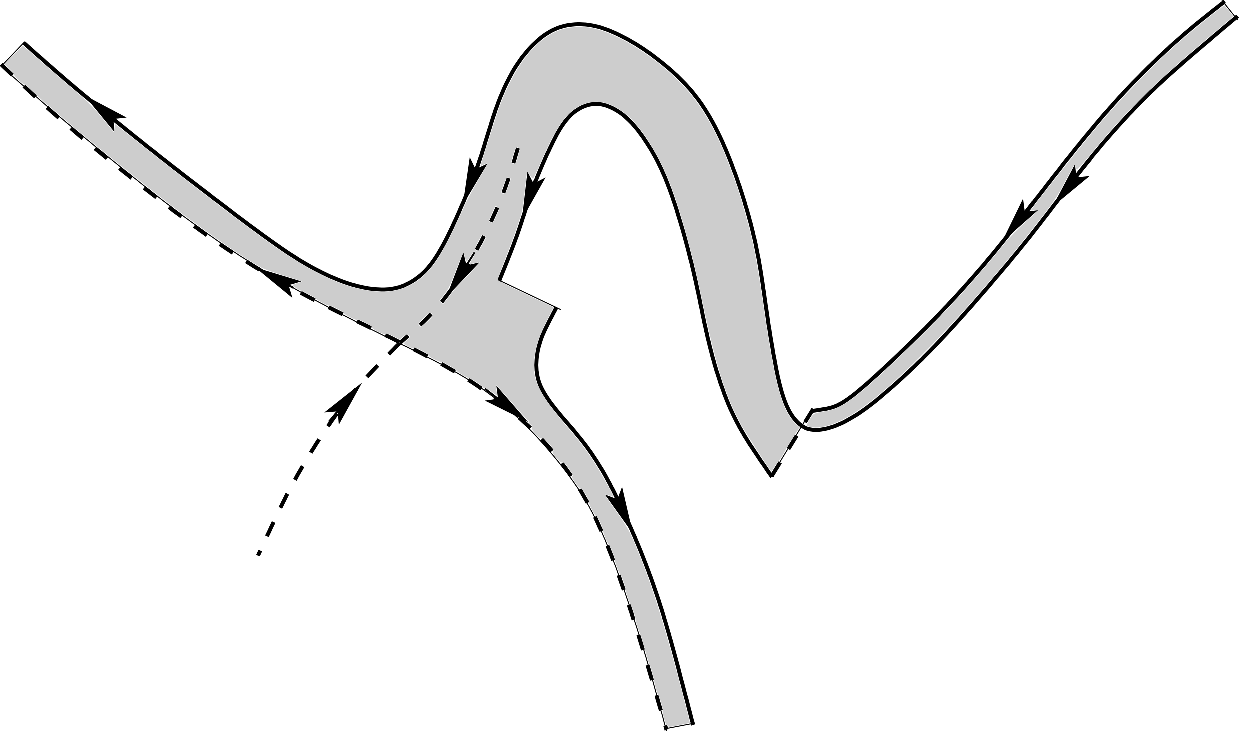}
\end{center}
\caption{Electron scattering at a saddle singular point of system 
(\ref{MFSyst}) caused by small-angle scattering on phonons}
\label{Fig10}
\end{figure}

 An electron trajectory (in $\, {\bf p}$ - space) almost does 
not change during small-angle scattering far from the saddle 
singular points of the system (\ref{MFSyst}). As can be seen, 
together with the original trajectory, it limits a narrow 
band on the Fermi surface, the width of which changes in each 
scattering event (Fig. \ref{Fig9}). If, at some point, a saddle 
singular point of the system (\ref{MFSyst}) falls inside this band, 
the initial and true trajectories of the electron quickly diverge 
in $\, {\bf p}$ - space (Fig. \ref {Fig10}). In the latter case, 
we can say that the electron was scattered at the saddle singular 
point of the system (\ref{MFSyst}), which was caused by its 
small-angle scattering on phonons. Scattering at singular points 
are independent and destroy electron velocity correlations at 
times exceeding $\, \tau_{0} (B, T)\, $.

\begin{figure}[t]
\begin{center}
\includegraphics[width=\linewidth]{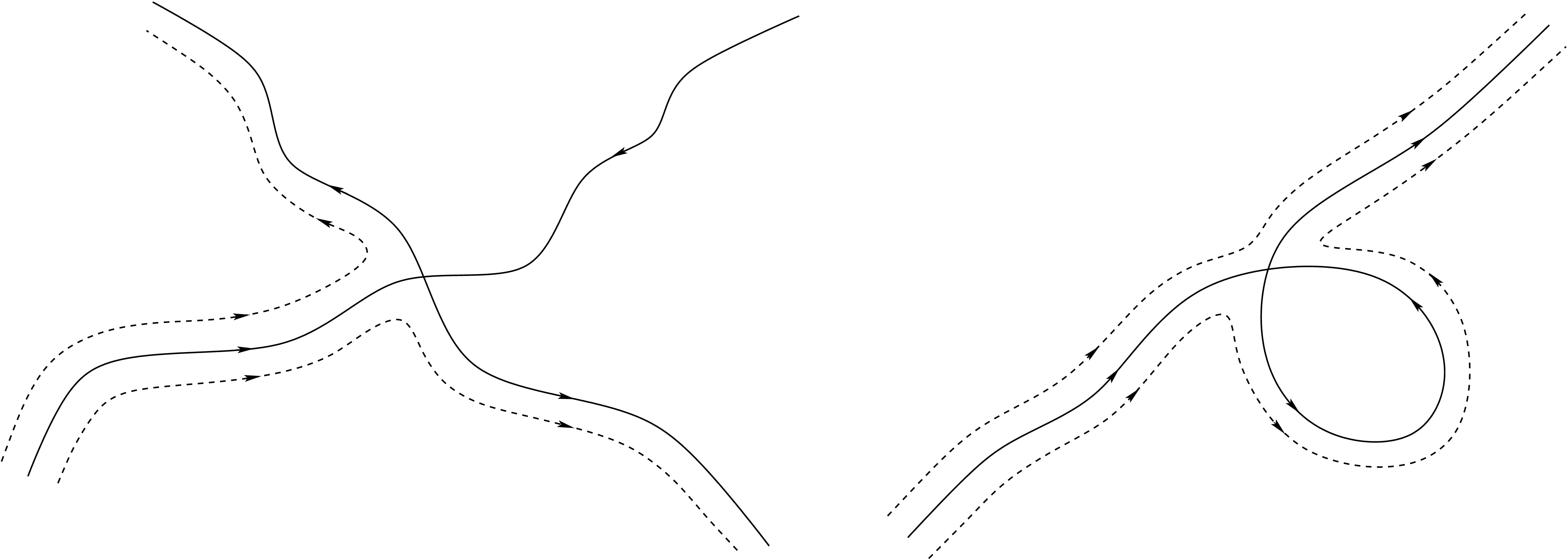}
\end{center}
\caption{Trajectories adjacent to the saddle singular points of 
system (\ref{MFSyst}) inside the surface 
$\, \widehat{S}_{F} ({\bf B}) \, $ and on its boundary.}
\label{Fig11}
\end{figure}

 Let us especially note here that the last property holds precisely 
for saddle points lying inside the carrier of open trajectories, 
and does not apply to singular points lying on the boundary of
$\, \widehat{S}_{F} ({\bf B}) \, $. This circumstance is due 
to the different geometry of the trajectories adjacent to such 
points in the first and second cases (Fig. \ref{Fig11}). As can be seen, 
reconstructions of open trajectories near points of the second type does 
not significantly change their geometry on large scales. Note also that 
the presence of singular points of the system (\ref{MFSyst}) inside 
the carrier of open trajectories is one of the main distinctive features 
of chaotic trajectories of the Dynnikov type and is not characteristic of 
trajectories of other types (see \cite{zorich1,dynn1,DynnBuDA,dynn2,dynn3}).

 To estimate the mean length $\, l_{0} \, $ corresponding to 
scattering at a singular point, we can equate the area 
$\, \Sigma (l_{0}) \, $ of the strip in Fig. \ref{Fig9} to the 
area of the carrier of chaotic trajectories (divided by the number 
of singular points inside it).

 The area $\, \widehat{S}_{F} ({\bf B}) \, $ is approximately 
equal to $\, p_{F}^{2} \, $ with a certain geometric coefficient, 
which can be noticeable greater than 1. On the surface 
$\, \widehat{S}_{F} ({\bf B}) \, $ of genus 3, however, there 
are 4 different saddle points of the system (\ref{MFSyst}), 
therefore for the corresponding area $\, \Sigma (l_{0}) \, $ 
we can write approximately
\begin{equation}
\label{SigmapF}
\Sigma (l_{0}) \,\,\, \simeq \,\,\, p_{F}^{2}
\end{equation}

 In particular, the condition 
$\, \tau_{0} (B, T) = \tau_{ph}^{\epsilon} (T) \, $ is determined 
by the inequality
$$\delta p_{0} \cdot \omega_{B} \, \tau_{ph}^{\epsilon} \, p_{F} 
\,\,\, \geq \,\,\, p_{F}^{2} \,\,\, , $$
i.e.
$$\omega_{B} \,\,\, \geq \,\,\, {k T \over \hbar} {T \over T_{D}} $$
(assuming that at length 
$\, l_{0} \simeq \omega_{B} \tau_{ph}^{\epsilon} p_{F} \, $ 
only one scattering by a phonon occurs).

\begin{figure}[t]
\begin{center}
\includegraphics[width=\linewidth]{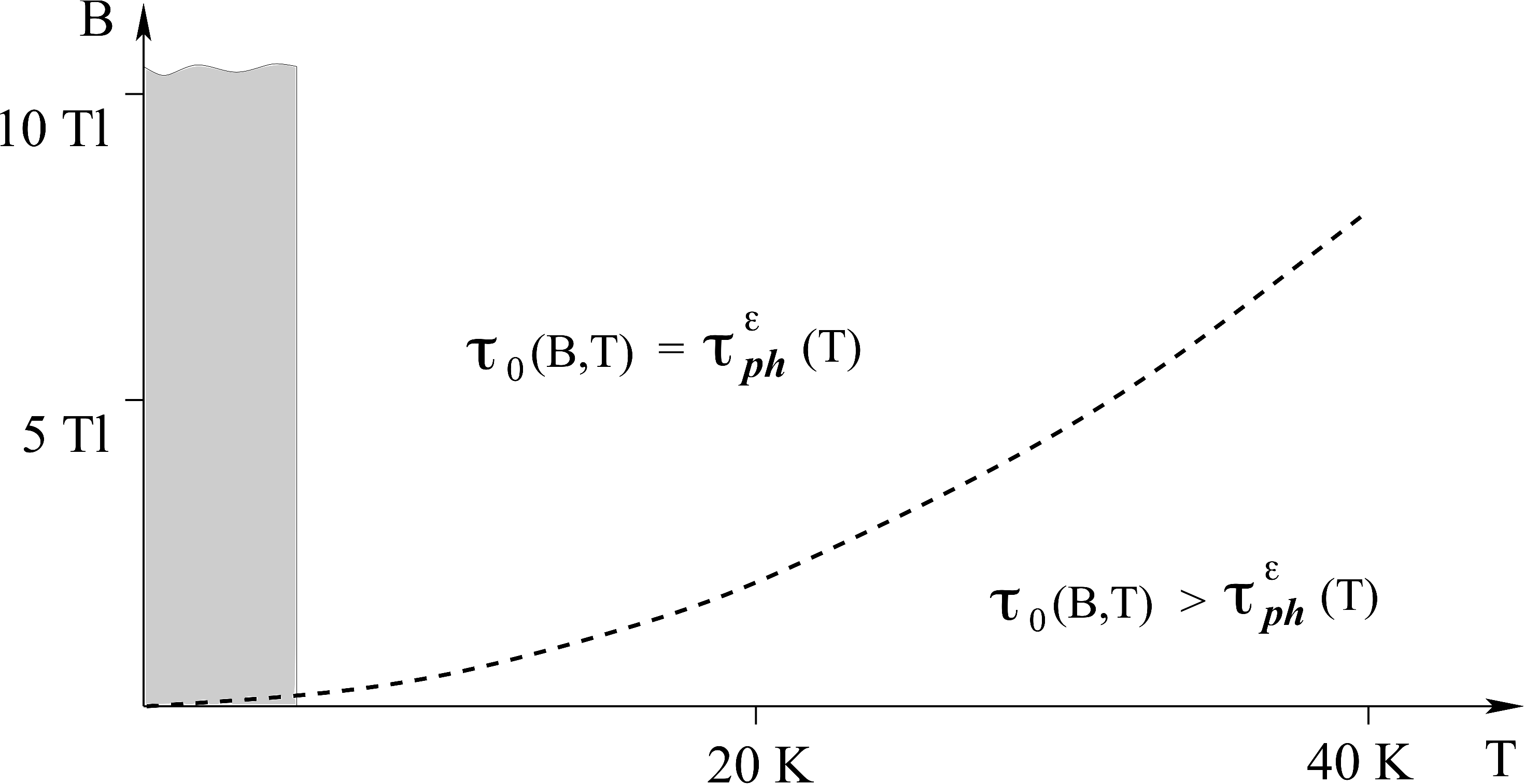}
\end{center}
\caption{Conditional curve separating areas with 
$\, \tau_{0} (B, T) = \tau_{ph}^{\epsilon} (T) \, $ and 
$\, \tau_{0} (B, T) > \tau_{ph}^{\epsilon} (T) \, $ for 
standard values of $\, \epsilon_{F} \, $ (5 {\rm eV}) 
and $\, T_{D} \, $ (300 {\rm K}). The shaded area corresponds 
to the values of $T$ at which the times $\, \tau_{ee} \, $ 
and $\, \tau_{imp} \, $ begin to play a major role 
(approximately).}
\label{Fig12}
\end{figure}

 The shape of the curve
\begin{equation}
\label{VidKrivoi}
\omega_{B} \,\,\, = \,\,\, {k T \over \hbar} {T \over T_{D}} \,\,\, ,
\end{equation}
separating the modes $\, \tau_{0} (B, T) = \tau_{ph}^{\epsilon} (T) \, $ 
and $\, \tau_{0} (B, T) > \tau_{ph}^{\epsilon} (T) \, $, for ``standard'' 
metal parameters is shown in Fig. \ref{Fig12}. It must be said, certainly, 
that this curve is to a large extent conditional, and we can rather talk 
about a certain region near it, separating the two indicated regimes. 
Its position also, in reality, strongly depends on the parameters of 
the conductor.

 The value $\, \omega_{B} \tau_{0} (B, T) \, $ on the curve 
(\ref{VidKrivoi}) is equal to
$$\omega_{B} \tau_{0} \,\,\, \simeq \,\,\, 
{k T \over \hbar} {T \over T_{D}} {\hbar \over k T} 
\left( {T_{D} \over T} \right)^{2} \,\,\, = \,\,\, {T_{D} \over T} $$
and satisfies the condition $\, \omega_{B} \tau_{0} \gg 1 \, $ in the 
interval of interest to us. This condition is also well satisfied 
(in the temperature range of interest to us) in the region 
$\, \tau_{0} (B, T) = \tau_{ph}^{\epsilon} (T) \, $ (above the curve).

 The fulfillment of the condition $\, \omega_{B} \tau_{0} \gg 1 \, $ 
in the area under the curve (\ref{VidKrivoi}) requires a separate study 
(primarily due to a decrease in the value of $\, \omega_{ B} \, $ in this 
area). In the general case, the dependence of time $\, \tau_{0} \, $ on the 
values of $B$ and $T$ here can be quite complex.

 In the limit $\, \tau_{0} (B, T) \gg \tau_{ph}^{\epsilon} (T) \, $ 
to determine the mean deviation $\, |\Delta p| \, $ along the 
corresponding section of the trajectory we can set
$$\left| \Delta p \right| \,\,\, \simeq \,\,\, \delta p_{0} \,
\sqrt{\tau_{0} \over \tau_{ph}^{\epsilon}} $$

 For the corresponding strip area in Fig. \ref{Fig9}, assuming 
$\, l_{0} \simeq \omega_{B} \tau_{0} p_{F} \, $, we can use then 
the estimate
$$\Sigma (\tau_{0}) \,\,\, \simeq \,\,\, \delta p_{0} \, 
\sqrt{\tau_{0} \over \tau_{ph}^{\epsilon}} \,\, 
\omega_{B} \, \tau_{0} \, p_{F} \,\,\, \simeq \,\,\, 
{T \over T_{D}} {\tau_{0}^{3/2} \over \sqrt{\tau_{ph}^{\epsilon}}} \,\,
\omega_{B} \, p_{F}^{2} $$

 Using the relation (\ref{SigmapF}), for the time 
$\, \tau_{0} (B, T) \, $ we obtain
\begin{equation}
\label{tau0} 
\tau_{0} (B, T) \, \simeq \, \left( 
{T_{D} \over T} {\sqrt{\tau_{ph}^{\epsilon}} \over \omega_{B}} 
\right)^{2/3} \simeq \, \left( {T_{D} \over T} \right)^{4/3} 
\left( {\hbar \omega_{B} \over k T} \right)^{1/3} {1 \over \omega_{B}} 
\end{equation}
and thus
$$\omega_{B} \, \tau_{0} (B, T) \,\,\, \simeq \,\,\,  
\left( {T_{D} \over T} \right)^{4/3} 
\left( {\hbar \omega_{B} \over k T} \right)^{1/3} $$ 

 It must be said, however, that the indicated limit, 
apparently, can be observed extremely rarely, and the 
dependence $\, \tau_{0} (B, T) \, $ rather has some 
intermediate form between (\ref{tau0}) 
and $\, \tau_{0} = \tau_{ph}^{\epsilon} \, $.

\vspace{1mm}

 Summarizing the above, we can see that when Dynnikov's 
chaotic trajectories appear on the Fermi surface, the behavior 
of the conductivity tensor in strong magnetic fields depends 
significantly on the value of $\, \tau_{ph}^{\epsilon} \, $ 
(or $\, \tau_ {0} (B, T)$). For ultrapure materials, moreover, 
it is possible to indicate temperature and magnetic field 
intervals where this dependence is decisive for the behavior 
of $\, \sigma^{kl} (B, T)\, $. In a more general situation, 
the relaxation time is also determined by the processes 
of electron-electron scattering and scattering by impurities, 
and is given by the relation
$$\tau^{-1} \,\,\, \simeq \,\,\, \tau_{0}^{-1} (B, T) 
\,\, + \,\, \tau_{ee}^{-1} (T) \,\, + \,\, \tau_{imp}^{-1} $$

 Let us note here that the above property is associated precisely 
with the contribution of chaotic trajectories to the tensor 
$\, \sigma^{kl} (B, T)\, $, in particular, in the accompanying 
contribution (\ref{Closed}) of closed trajectories the 
time $\tau$ is determined by the relation
$$\tau^{-1} \,\,\, \simeq \,\,\, 
\tau_{ee}^{-1} (T) \,\, + \,\, \tau_{imp}^{-1} $$
(considering the time $\, \tau_{ph}^{p} \, $ noticeably larger 
in our temperature range).

 In the limit of very low temperatures ($T < 1\, {\rm K}$), 
where the relations 
$$\tau_{ph}^{\epsilon}, \, \tau_{ee} \,\,\, \gg \,\,\, \tau_{imp} $$
hold, the time $\, \tau_{imp}\, $ plays the role of the universal 
relaxation time when calculating the conductivity tensor.

 As we have already noted, due to the specificity of the 
contribution of chaotic trajectories to $\, \sigma^{kl} (B, T)\, $, 
its dependence on $\tau$ is most pronounced when the scaling 
parameters $\, \alpha_{l} \, $ are noticeably different from $1/2$. 
This situation, as a rule, also corresponds to the greatest 
anisotropy of chaotic trajectories in planes orthogonal to ${\bf B}$ 
(for example, $\alpha_{1} < 1/2$, $\alpha_{2} > 1/2$). When 
determining the dependence $\, \tau_{0} (B, T) \, $ in the 
interval considered above, one can use the values of 
$\, \alpha_{l} \, $ measured from the dependence 
$\, \sigma^{kl} (B) \, $ in the limit of very low temperatures 
($\tau = \tau_{imp}$), where they can be determined with the 
greatest accuracy.

\section{Conclusion}
\setcounter{equation}{0}

 The paper examines the behavior of the magnetic conductivity 
of a metal in a special situation, namely, when chaotic electron 
trajectories arise on the Fermi surface. It is shown that, 
in a certain range of temperatures and magnetic fields, 
the behavior of conductivity in this case is determined by 
the electron-phonon energy relaxation time 
$\, \tau_{ph}^{\epsilon} \, $ (or the associated time 
$\, \tau_{ 0} (B, T)$), which usually does not play a role 
in calculating conductivity. The reason for this is the 
scattering of electrons at singular points of the system, 
which describes the dynamics of an electron on the 
Fermi surface in the presence of an external magnetic field. 
Such scattering is actually caused by small-angle scattering 
of electrons on low-momentum phonons and provides rapid momentum 
relaxation of electrons in this situation. The general dependence 
of the conductivity tensor on the values of $T$ and $B$ is 
determined both by the function $\, \tau_{0} (B, T) \, $ 
and by the geometric features of chaotic trajectories.

\end{document}